\documentclass{article}[12pt]
\usepackage{graphicx}
\usepackage{caption}
\usepackage{subcaption}
\usepackage[utf8]{inputenc}
\usepackage[toc,page]{appendix}
\usepackage{amsmath}
\usepackage{csquotes}
\usepackage{gensymb}
\usepackage{multicol}
\usepackage{a4wide, graphics, amsmath, amssymb, cite,nicefrac}
\usepackage[verbose]{wrapfig}
\usepackage{authblk,hyperref}
\usepackage{url,amsfonts}
\usepackage{physics}
\usepackage{xcolor}
\hypersetup{colorlinks=true, linkcolor=blue, citecolor=red, linktoc=page}
\begin{document}
	\begin{titlepage}

		\vskip 1.0cm

	\begin{center}
			{\boldmath \Huge{Page Curve and Island in EGB gravity}}

		\vskip 2.0cm
		{\bf \large Ankit Anand \footnote{E-mail - Anand@physics.iitm.ac.in}  }
		
		\vskip 0.5cm
		{\it ${}^1$%
			Department of Physics, \\
			Indian Institute of
			Technology
			Madras, Chennai
			600 036, India}\\
		
		\vskip 10pt
			\end{center}
		
		\vskip 1.5cm
		
		\begin{center}
			
			 {\bf ABSTRACT}
		
			 \end{center}%
			 
		 In this paper, the information paradox on lower-dimensional Gauss–Bonnet gravity, known as a three-dimensional EGB black hole, is studied using the quantum extremal island approach. For that, we connect an auxiliary flat bath system to this timelike singularity spacetime and estimate the entropy of hawking radiation in its asymptotic regions when gravity is weak. The addition of island areas to the entanglement wedge of radiation causes its entropy to obey the Page curve, as shown. The quantum extremal surface(QES), or island boundary, is located outside the horizon. This yields a time-independent equation for Hawking radiation fine-grained entropy that is compatible with the appropriate Page curve. We also discover the modifications to this entropy and Page time.

	\end{titlepage}

\section{Introduction}
From the quantum gravity theory point of view, the investigation of the information loss paradox has been one of the most important challenges \cite{Hawking:1975vcx}. Hawking radiation has the same properties as thermal radiation, implying that the entanglement entropy(EE) outside the black hole(BH) is increasing monotonically. Quantum mechanics, on the other hand, mandates that the EE at the end of the evaporation be zero because it must be the pure state. The EE changes with time and results in the generation of the so-called Page curve \cite{Hawking:1976ra,Page:1993wv}. As a result, the information loss paradox is transformed into getting the Page curve for Hawking radiation's EE. By incorporating a scale for the time measurement known as the Page time $t_p$, the Page curve effectively  helps to answer whether the problem of the information loss paradox can be resolved or not. In terms of the Page curve, the information loss paradox can be explained as follows. The quantum field's von Neumann entropy on the radiation region $R$ outside the BH gives the fine-grained entropy of Hawking radiation. We assume that the entire Cauchy slice is in the pure state, the fine-grained entropy of radiation $S(R) = S(R_c )$, where $S(R_c)$ is the BH subsystem's fine-grained entropy. Because the whole system is the BH Plus radiation region, the region $R_c$ should correspond to the BH's fine-grained entropy,  and the region $R$ corresponds to the radiation's fine-grained entropy. This means that there is always a condition that $S(R)$ has to satisfy i.e., $S(R)\leq S_{BH}$, where $S_{BH}$ is the BH subsystem's coarse-grained entropy. However, it has been discovered that $S(R) > S_{BH}$ immediately after the Page time $t_p$, resulting in the contradiction.

\par
The Page curve has recently been postulated to develop from the action of islands \cite{Penington:2019npb,Almheiri:2019psf,Almheiri:2019hni,Almheiri:2019yqk}. The density matrix of R is generally determined by taking the partial trace across the states in R$_{c}$, which is the complementary area of $R$, when considering the state of the Hawking radiation as that in a region R outside the BH. Recent research \cite{Almheiri:2019qdq,Chen:2019uhq,Hashimoto:2020cas,Anegawa:2020ezn,Matsuo:2020ypv,Akal:2020twv,Almheiri:2020cfm,Hartman:2020swn,Dong:2020uxp,Balasubramanian:2020xqf,Raju:2020smc,Alishahiha:2020qza,Saha:2021ohr,Yu:2021rfg,Azarnia:2022kmp,Yadav:2022fmo,Ahn:2021chg,Basu:2022reu,Afrasiar:2022ebi,Omidi:2021opl} has revealed that specific supplementary regions known as islands contribute to the entropy of Hawking radiation, with their boundaries defined as surfaces known as QES. This demonstrates the existence of a non-trivial QES in spacetime at Page time, which cancels out the time-dependency of $S(R)$ and leads to a saturated Hawking radiation's fine-grained entropy. By inclusion of island contributions, Hawking's fine-grained entropy is
\cite{Almheiri:2019psf}
\begin{equation} \label{S(R)}
    S(R) = \text{min} \;\;\;\;\; \text{ext} \left[\frac{Area(\partial I)}{4G_N}+ S_{\text{matter}}(I \cup R)\right] \ .
\end{equation}

\par

The island rule was first postulated due to a conjectured QES prescription, and it was recently deduced using the replica approach for the gravitational path integral. When the replica trick \cite{Callan:1994py,Holzhey:1994we,Calabrese:2009qy} is used in gravitational theories, only the boundary conditions of the replica geometries can be fixed, and new saddles, where bulk wormholes join distinct copies of spacetime, must be considered. Replica wormholes, as these new saddles are known, lead to islands \cite{Penington:2019kki,Almheiri:2019qdq}.
When using replicas and in the semi-classical gravity limit, there is of the partition function of the geometry is the one with the lowest entanglement entropy dominant. The replica trick for gravitational theories yield the same formula \eqref{S(R)} as the QES prescription in this fashion.
\par
Higher-curvature gravity theories continue to pique attention, partly because the majority of quantum gravity theories indicate that such corrections modify the Einstein–Hilbert(EH) action and because it provides a unique setting to understand classical gravity in strong Gravitational fields. Lovelock theories \cite{Lovelock:1971yv}, which are the broadest theories formed from the Riemann curvature tensor and preserve metric's second-order motion equations, are the most promising and most investigated options. For $D > 5$, although they have the feature that any additional contributions they make to the action are topological or zero. The cosmological and Einstein–Hilbert factors are combined in this theory, which includes new corrections for each odd spacetime dimension above four. The Gauss-Bonnet term is the first new correction of this type 
\begin{equation} \label{GB term}
\mathcal{G} = R_{\alpha \beta \gamma \tau} R^{\alpha \beta \gamma \tau} - 4  R_{\alpha \beta} R^{\alpha \beta} + R^2 \ ,
\end{equation}
has five or more dimensions of activity. Significantly, for D $<$ 5, the new terms that come from this are either zero, or it is topological, establishing Einstein's theory as the most general second-order metric theory of gravity in four dimensions. Recently, a novel approach \cite{Glavan:2019inb} for overcoming this constraint for Gauss-Bonnet gravity (the most fundamental of the Lovelock theories) has attracted attention. By considering the theory's parameter as the spacetime dimension and scaling the Gauss-Bonnet coupling ($\alpha$), it is possible to build $D = 4$ and $D = 3$ variations of this theory. Even though the initial approach involved taking limits of solutions to the field equations, which raised several consistency concerns \cite{Gurses:2020ofy,Hennigar:2020lsl,Bonifacio:2020vbk}, maintaining consistency in $D < 5$ can be obtained unbiased of any real solutions or additional dimensions \cite{Fernandes:2020nbq,Hennigar:2020fkv}. This approach is a modification of one previously in use to reach a D $\rightarrow 2$ limit of GR \cite{Mann:1992ar}, and it is congruent with a newly proposed dimensional reduction technique \cite{Lu:2020iav,Kobayashi:2020wqy}, the inside space must be flat. The resulting scalar-tensor theory of gravity is as follows
\begin{equation} \label{Action}
    S= \int d^Dx \sqrt{-g} \left[R - 2 \Lambda+\alpha \left( \phi \mathcal{G} + 4G^{\alpha \beta} \partial_\alpha\phi \partial_\beta \phi - 4 (\partial \phi)^2 \Box \phi +2 ((\nabla \phi)^2)^2 \right)\right] \ ,
\end{equation}
where $\mathcal{G}$ is Gauss-Bonet term given in equation \eqref{GB term} and $G_{\alpha \beta}$ is  the Einstein tensor. The Gauss–Bonnet term vanishes identically for the case of $D<4$ dimensions. For $D=3$, no scalar degree of freedom is propagating \cite{Lu:2020mjp}. The repercussions of lower-dimensional Gauss–Bonnet gravity, including BHs, solutions that radiate and collapse, BH thermodynamics, and additional physical effects,  have been studied extensively. Till now, only the spherically symmetric BH has been studied.
\par
In general, the entropy formula for higher derivative gravity may be written as 
\begin{equation} \label{Stotal}
    S_{\text{total}} = S_{\text{gravity}} + S_{\text{matter}} \ , 
\end{equation}
where $S_{\text{gravity}}$ and $S_{\text{matter}}$ are the gravitational and matter contributions to total entanglement entropy, respectively. In higher derivative gravity theories, the first component in equation \eqref{Stotal} may be determined using the Dong formula \cite{Dong:2013qoa} while the $S_{\text{matter}}$ term can be calculated using the Cardy formula \cite{Calabrese:2005zw,Calabrese:2009ez}. Then we must extremize the overall entanglement entropy with respect to the location of the island surfaces. If there are several surfaces, we must select the surface with the smallest area from among those surfaces. The authors of \cite{Alishahiha:2020qza}constructed Page curves of Schwarzschild black holes in the presence of higher derivative components that are $\order{R^2}$. Following \cite{Alishahiha:2020qza}, \cite{Yadav:2022fmo} calculates the Page curves of an eternal Reissner-Nordström black hole in four dimensions in the presence of $\order{R^2}$ terms as discussed in \cite{Alishahiha:2020qza} and in Einstein-Gauss-Bonnet gravity \cite{Fernandes:2020rpa}.
Throughout the paper we set $l=1$.

This paper is organized as follows. In Section \ref{BTZ}, we set the stage for further discussions (see the reference\cite{Saha:2021ohr}). In Section \ref{EGBdi}, we discuss the EGB gravity. In section \ref{Islanddi} we go over how to compute the entanglement entropy without and with the island in EGB gravity. Section \ref{Page time and scrambling time} displays the Page curve and discusses the Page time and scrambling time. Section \ref{Conclusion} is devoted to conclusions.

\section{Preliminaries} \label{BTZ}
\subsection{BTZ Black Hole} 
Let's start with the metric of BTZ BH  
\begin{equation} \label{BTZ bh}
    dS^2 = -\left(r^2-r_+^2\right) dt^2 + \left(\frac{1}{r^2-r_+^2}\right)dr^2+r^2 d\phi^2 \ ,
\end{equation}
where $r_+$ represents the location of the horizon. The thermodynamics of BTZ well studied in literature and can be written as 
\begin{eqnarray}
 M &=& \frac{m}{8} = \frac{r_+^2}{8}  \ , \hspace{0.5cm} T = \frac{f'}{4 \pi} = \frac{r_+}{2 \pi } \ , \hspace{0.5cm}  \Omega =0 \ , \nonumber \\
 S &=& \frac{\pi r_+}{2}  \ , \hspace{0.5cm} P=\frac{1}{8 \pi } \ , \hspace{0.5cm} V=\pi r_+^2 \ . 
\end{eqnarray}
The Kruskal extension of BTZ spacetime with identification as 
\begin{center}
    Right Wedge : $U=-e^{-k(t-r_*)}$ \;\;\;\;\; and \;\;\;$V=e^{k(t-r_*)}$ \\
   \;\;\;\;\; Left  Wedge \;\;: $U=e^{-k(t-r_*)}$ \;\;\;\;\; and \;\;\; $V=-e^{k(t-r_*)}$
\end{center}
where the surface gravity is defined as $k = \frac{2 \pi}{\beta}$, and the tortoise coordinate is indicated by $r_*$ and given as 
\begin{equation}
    r_*=\int \frac{dr}{f(r)} = \frac{1}{2r_+}\left[\text{log}\left(\frac{|r-r_+|}{r+r_+}\right) \right] \ .
\end{equation}
Using the linking auxiliary baths \cite{Rocha:2008fe}\cite{Yu:2021rfg}, we can now make the BTZ spacetime boundary transparent. We can treat the bath as a flat Minkowski spacetime devoid of gravitational effects and suppose it is in thermal equilibrium with the black hole. In terms of Kruskal coordinate, the metric can be written as
\begin{equation}
     dS^2= \Omega^{-2} dU dV + r^2 d\phi^2 ,
 \end{equation}
 with $\Omega=\frac{r_+}{r+r_+}$ and $\phi$=constant represents the two dimensional space. By ignoring the island's contribution, eq.\eqref{S(R)} takes the conventional form $S(R)=S_{gen}(R)$, meaning that the von Neumann entropy of quantum fields must be computed on the union of $R_+$ and $R_-$.

 \subsection{ Island and the Information Recovery} 
 We start by calculating the radiation's EE in the absence of islands. The area term does not affect the gravitational fine-grained entropy \eqref{S(R)} in this case. Only the right and left wedges of the Penrose diagram have Hawking radiation zones, and the borders of these entanglement regions are marked by $b_+$ and $b_-$, respectively
	\begin{figure}[ht]
\begin{subfigure}{.5\textwidth}
  \centering
  \includegraphics[width=.8\linewidth]{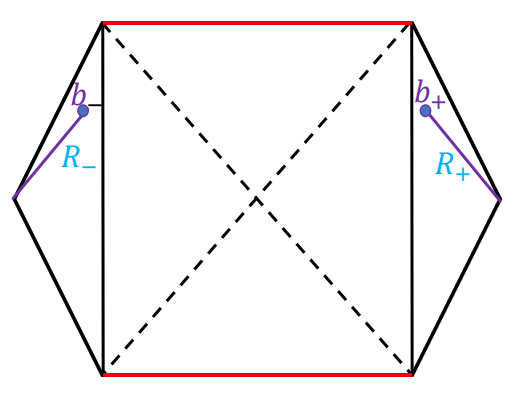}  
  \label{fig:sub-first}
\end{subfigure}
\begin{subfigure}{.5\textwidth}
  \centering
  \includegraphics[width=.8\linewidth]{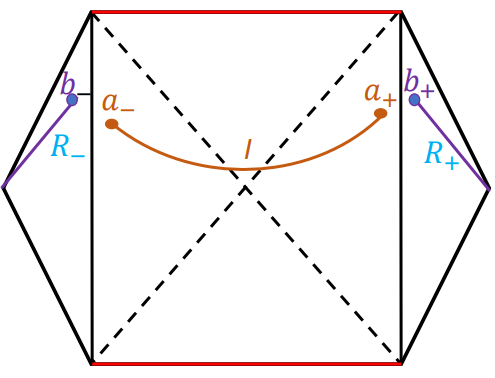}  
  \label{fig:sub-second}
\end{subfigure}
\caption{Penrose diagrams of the BTZ BH (left) and one with an island $I$ (right)}
\label{Ads4 2}
\end{figure}	
As a result, the generalized EE is made up entirely considering the von Neumann entropy of the constrained quantum matter proposed by
\begin{equation}
    S_{\text{gen}}= \frac{c}{3} \text{Log}[\ell(b_+,b_-)] \ ,
\end{equation}
where $\ell(b_+,b_-)$ denotes the distance between $b_+$ and $b_-$. The expression for the distances is already calculated in \cite{Saha:2021ohr}. So we have 
\begin{equation}
    S_{\text{gen}}= \frac{c}{6} \text{Log}\left[ \frac{4(b^2-r_+^2)}{r_+^2}     \text{Cosh}^2(2r_+t_b)\right] \ ,
\end{equation}
$c$ is the central charge. At late time i.e., $t \rightarrow \infty$ then $\text{Cosh}(r_+t)\sim e^{r_+t}$, with this approximation the above expression reduces to  
\begin{equation}\label{Srbtz}
    S(R) \sim \frac{c}{3} r_+ t \ ,
\end{equation}
with the identification of $t_b$ as $t$. We can see from equation \eqref{Srbtz}, that in the absence of an island surface, the EE of the Hawking radiation increases linearly with time and becomes infinite at late times, resulting in BTZ BH's information paradox.

\par
Now we will show that at the late time( i.e., after page time), an island appears and result in the constant value of EE of the Hawking radiation.

\par
To check whether the island is present or not in the BTZ+Bath system, for the s-wave approximation to be effective, we presume that the right wedge is far apart from the left wedge. Moreover, the von Neumann entropy of quantum matter is adequate for unifying the radiation and island over the whole region.

\begin{equation}\label{Sr with island}
    S (R \cup I) = \frac{c}{6}\text{log} \left[\frac{\ell(a_+,a_-)\ell(b_+,b_-)\ell(a_+,b_+)\ell(a_-,b_-)}{\ell(a_+,b_-)\ell(a_-,b_+}\right] \ ,
\end{equation}
where $a_\pm = (\pm t_a,a)$. Because large distances between two wedges are assumed, it follows that
\begin{equation}
    \ell(a_+, a_-) \simeq \ell(b_+, b_-) \simeq \ell(a_\pm, b_\mp) \gg \ell(a_\pm, b_\pm) \ .
\end{equation}

So, the EE of the whole system can be written as the sum of the $R \cup I$ in left as well as right wedges.

\par
The expression of the generalized entropy will be
\begin{eqnarray} \label{Sgenisland}
    S_{\text{gen}} &=& \frac{\pi a}{G_N} + \frac{c}{3}\text{log}[\ell(a_+,b_+)] \nonumber  \ , \\
    &=& \frac{\pi a}{G_N} + \frac{c}{6} \text{log} \Bigg[\frac{(a+r_+)(b+r_+)}{r_+} \Bigg[ \left(\frac{a-r_+}{a+r_+}\right) + \left(\frac{b-r_+}{b+r_+}\right)  \nonumber \\
 && - 2 \Bigg(\sqrt{ \left(\frac{b-r_+}{b+r_+}\right)} \sqrt{\left(\frac{a-r_+}{a+r_+}\right) } \text{Cosh}(r_+(t_a-t_b))  \Bigg]  \ .
\end{eqnarray}
Now, Extremizing the $S_\text{gen}$ \eqref{Sgenisland} with respect to $t_a$ and equating to zero results in $t_a =t_b$. Substituting this value to the equation \eqref{Sgenisland} and then by extremizing w.r.t $a$ i.e., $\frac{\partial S_{\text{gen}}}{\partial  a}=0$ reveals that
\begin{equation} \label{islandbtz}
    a = r_+ + \frac{1}{2} \left(\frac{cG_N}{6\pi}\right)^2   + \dots \ .
\end{equation}
Substituting this value in  $S_\text{gen}$ we get
\begin{eqnarray} \label{Sgenfinal}
     S_\text{gen} &=& \frac{\pi r_+}{G_N} +\frac{c^2}{72} \frac{G_N}{\pi } \nonumber \\
     &=& 2 S_{BH} +\cdots  \ .
\end{eqnarray}
The above phrase contains a few distinguishing features. The statement includes universal corrections to the BH's Hawking entropy, in addition to the leading item $S_{BH}$. Because these adjustments are quantum gravity signals, this is a pleasant surprise. It is astounding how the mutual information interpretation explains how the aforementioned conclusion yields the right Page curve, and so resolves the information loss paradox.
\subsection{Page Curve}
EE increases linearly with time in the absence of an island surface, according to equation \eqref{Srbtz}, and is constant at late times, according to equation \eqref{Sgenfinal}. The Page time island surface forms, saturating the linear expansion of entanglement entropy and reaching a constant value double the Bekenstein-Hawking entropy of the BTZ BH, and we have the Page curve of the BTZ BH.
\begin{itemize}
    \item \textbf{Page time $(t_p)$ :} Page time is defined as the point at which the Hawking radiation's EE begins to decline to zero for an evaporating black hole and achieves a constant value for an eternal BH,
    \begin{eqnarray}
        t_p &=& \frac{6 S_{BH}}{c r_+} + \cdots \nonumber \\
           &=& \left(\frac{6 \beta}{ c \pi}\right)S_{BH} -  \cdots 
    \end{eqnarray} 
    So, the leading component in the above equation represents the page time($t_p$).

    \item \textbf{Scrambling time:  $(t_\text{scr})$ :} The time interval during which we retrieve the information sent into the BH in the form of Hawking radiation is known as scrambling time. Furthermore, it is the amount of time taken for the information to reach the island's surface. If we wish to transfer data from the cutoff surface $r = b$ to the BH, the time it takes to get to the island surface $r = a$ is: 
    \begin{eqnarray}
     t_{\text{scr}} &\equiv& |r_*(b) -r_*(a)| \nonumber \\
                    &=& \frac{1}{2r_+} \text{log}\left(\frac{b-r_+}{b+r_+}\frac{a+r_+}{a-r_+}\right) \nonumber \\
                    &\simeq& \frac{\beta}{2 \pi}\text{log}\left(\frac{r_+}{c G_N}\right) + \cdots \nonumber \ .
    \end{eqnarray}
   In the above calculation, we have set the order of $b$ as same as that of $r_+$, fix the island location as in equation \eqref{islandbtz}, and by assuming the $S_{\text{BH}} \simeq \frac{r_+}{G_N}$. We have an expression for Scrambling time as 
   \begin{equation}
        t_{\text{scr}} \simeq \frac{\beta}{2 \pi}\text{log}\left(S_{\text{BH}}\right)
   \end{equation}
\end{itemize}
\section{EGB Gravity in Lower dimensions} \label{EGBdi}

Let's start with the action that describes the Gauss-Bonnet (GB) gravity in the lower dimension from eq.\eqref{Action} and is given by
\begin{equation} \label{Action(2+1)}
     S= \int d^3x \sqrt{-g} \left[R - 2 \Lambda+\alpha \left( \phi \mathcal{G} + 4G^{\alpha \beta} \partial_\alpha\phi \partial_\beta \phi - 4 (\partial \phi)^2 \Box \phi +2 ((\nabla \phi)^2)^2 \right)\right] \ ,
\end{equation}
with $\mathcal{G}=R_{\alpha \beta \gamma \delta}R^{\alpha \beta \gamma \delta}-R_{\alpha \beta}R^{\alpha \beta}+R^2$, in addition having extra scalar field $\phi$ and the GB coupling $\alpha$. Following \cite{Cuadros-Melgar:2022lrf,Hennigar:2020fkv}, some extra term has to be added in action to reduce the complications, and the term is 
\begin{equation} \label{Extra Action}
    S_\lambda = -2 \int d^3x \sqrt{-g}\left[ \lambda e^{-2 \phi}(R+6(\partial \phi)^2)+3 \lambda^2 e^{-4 \phi}\right] \ ,
\end{equation}
where the interior space's curvature is represented by the $\lambda$. We investigate the equations of motion resulting from the action (1) along with the extra term (2) and the line element for deriving BH solutions to the GB gravity; we start with the line element
\begin{equation}
    ds^2 = -f(r)dt^2 + \frac{1}{h(r)f(r)}dr^2 + r^2\left(d\phi^2-\frac{J^2}{2r^2}dt\right)^2 \ ,
\end{equation}
where $J$ is some constant. We have three radial coordinate dependence functions $f(r),h(r)$ and the scalar field $\phi(r)$. So, varying the action (eq. \eqref{Action(2+1)} along with eq.\eqref{Extra Action}), we have three different equations of motion. As we are interested in BTZ-like solutions, we set $h(r)=1$. From \cite{Cuadros-Melgar:2022lrf,Hennigar:2020fkv}, 
\begin{enumerate}
    \item For the case \enquote{$\phi$=costant} and \enquote{$\lambda=0$} :  \\
    
    This choice results in 
    \begin{equation}
       f(r)=-m+\frac{r^2}{l^2}+\frac{J^2}{4r^2} \;\;\;\;\;\; \Lambda=-\frac{1}{l^2} \ ,
    \end{equation}
    where $m$ is BH mass  and $J$ is  the angular momentum of the BH. This situation is nothing but the BTZ case. 

    \item For the case \enquote{$\phi(r)=\text{ln}\left(\frac{r}{l}\right)$} and \enquote{$\lambda=0$} :  \\ 
    
    This situation results in 
    \begin{equation} \label{f(r)}
        f_\pm = -\frac{r^2}{2 \alpha}\left(1\pm \sqrt{1+\frac{4\alpha}{r^2}\left[\frac{r^2}{l^2}-m\right]}\right) \ ,
    \end{equation}

    As the GB coupling constant approaches $0$, the $f(r)_+$ of solution \eqref{f(r)} lacks a well-defined limit, i.e., in the limit $\alpha\rightarrow 0$, $f(r)_+$ blows up to infinity. So, only the $f(r)_-$ has a well-defined limit as $\alpha\rightarrow 0$, and it gives the BTZ metric. The $f(r)_-$ in this limit is 
    \begin{equation}
        f(r)_-=\frac{r^2}{l^2}-m-\frac{\alpha}{r^2}\left(\frac{r^2}{l^2}-m\right)^2 + \order{\alpha^2} \ .
    \end{equation}
    
\end{enumerate}
 Now, the authors of \cite{Hennigar:2020fkv} discussed the entropy expression based on the Iyer-Wald method \cite{Wald:1993nt}\cite{Iyer:1994ys}, expression for entropy is 
 \begin{equation}
     S=\frac{\pi r_+}{2} \left\{1-2 \lambda \alpha e^{-2 \phi}\right\} \ .
 \end{equation}

  As we are working in $\lambda=0$ and $\phi(r)=\text{ln}\left(\frac{r}{l}\right)$. the thermodynamics of such BHs can be written as 
\begin{eqnarray}
 M &=& \frac{m}{8} = \frac{r_+^2}{8l^2}  \ , \hspace{0.5cm} T = \frac{f'}{4 \pi} = \frac{r_+}{2 \pi l^2} \ , \hspace{0.5cm}  \Phi =0 \ , \nonumber \\
 S &=& \frac{\pi r_+}{2}  \ , \hspace{0.5cm} P=\frac{1}{8 \pi l^2} \ , \hspace{0.5cm} V=\pi r_+^2 \ .
\end{eqnarray}
It is the same as the well-known BTZ black hole in Einstein's gravity. The first law of BH thermodynamics states that $\delta M = T \delta S + V \delta P + \Phi_\alpha \delta  $, it is to verify that it is verified with above parameters. Also the Smarr formula $0 = T S- 2P V + 2 \Phi_\alpha \alpha $, is also verified. This is interesting because, regardless of the value of $\alpha$, the thermodynamic parameters remain the same even if the curvature is not constant.
\par 

The metric is 
\begin{equation}
    dS^2 = -f(r)dt^2 + \frac{1}{f(r)}dr^2+r^2 d\phi^2 \ ,
\end{equation}
where $f(r)=\frac{r^2}{l^2}-m-\frac{\alpha}{r^2}\left(\frac{r^2}{l^2}-m\right)^2 + \order{\alpha^2} = \frac{r^2-r_+^2}{l^2}-\frac{\alpha}{r^2}\left(\frac{r^2-r_+^2}{l^2}\right)^2 + \order{\alpha^2}$. In the limit of small $\alpha$, it reproduces the typical BTZ solution. $r_+$ is the location of the horizon, and it matches with the BTZ horizon of the same mass.
 
\par 
Working with the two-dimensional surface with $\phi$ = constant, the metric can be written as 
\begin{equation}
    dS^2 = -f(r)dt^2 + \frac{1}{f(r)}dr^2 \ .
\end{equation}
Furthermore, the Kruskal coordinate can be defined to achieve the maximal BTZ spacetime extension. \\
\begin{center}
    Right Wedge : $U=-e^{-k(t-r_*)}$ \;\;\; and 
 \;\;\;\;\;$V=e^{k(t-r_*)}$ \\
    \:\;Left Wedge \;\;: $U=e^{-k(t-r_*)}$ \;\;\;\;\;\;\; and \;\;\;\;\; $V=-e^{k(t-r_*)}$
\end{center}
with the expression of surface gravity is $ k = \frac{2 \pi}{\beta}$ and $r_*$ is the tortoise coordinate and given by 
\begin{equation}
    r_*=\int \frac{dr}{f(r)} = \frac{1}{2r_+}\left[\text{log}\left(\frac{|r-r_+|}{r+r_+}\right) - \gamma \hspace{0.1cm} \text{log}\left(\frac{| r-\gamma r_+|}{ r+ \gamma r_+}\right)\right] \ ,
\end{equation}
where $ \gamma = \sqrt{\frac{\alpha}{1-\alpha}}$.
 As we discussed above, the procedure of linking flat space. In terms of Kruskal coordinate, the metric can be written as 
 \begin{equation}
     dS^2= \Omega^{-2} dU dV \ ,
 \end{equation}
 with $\Omega=\frac{r+r_+}{rr_+}\sqrt{\left(\left[\frac{| r-\gamma r_+|}{r+\gamma r_+}\right]^\gamma [r^2+(r^2-1)\gamma^2] \right)}$ . $R_-$ and $R_+$ are the left and right wedges of the radiation regions, $b_-$ and $b_+$ are the boundaries of the $R_-$ and $R_+$, and $a_-$ and $a_+$ are the boundaries of the island surface in the left and right wedges for an eternal EGB black hole.

\section{Island and the Information Recovery} \label{Islanddi}
\subsection{Without Island}
In this subsection, we discuss the calculation of EE without an island.
\begin{figure}[ht]
	\begin{center}
		\includegraphics[scale=0.45]{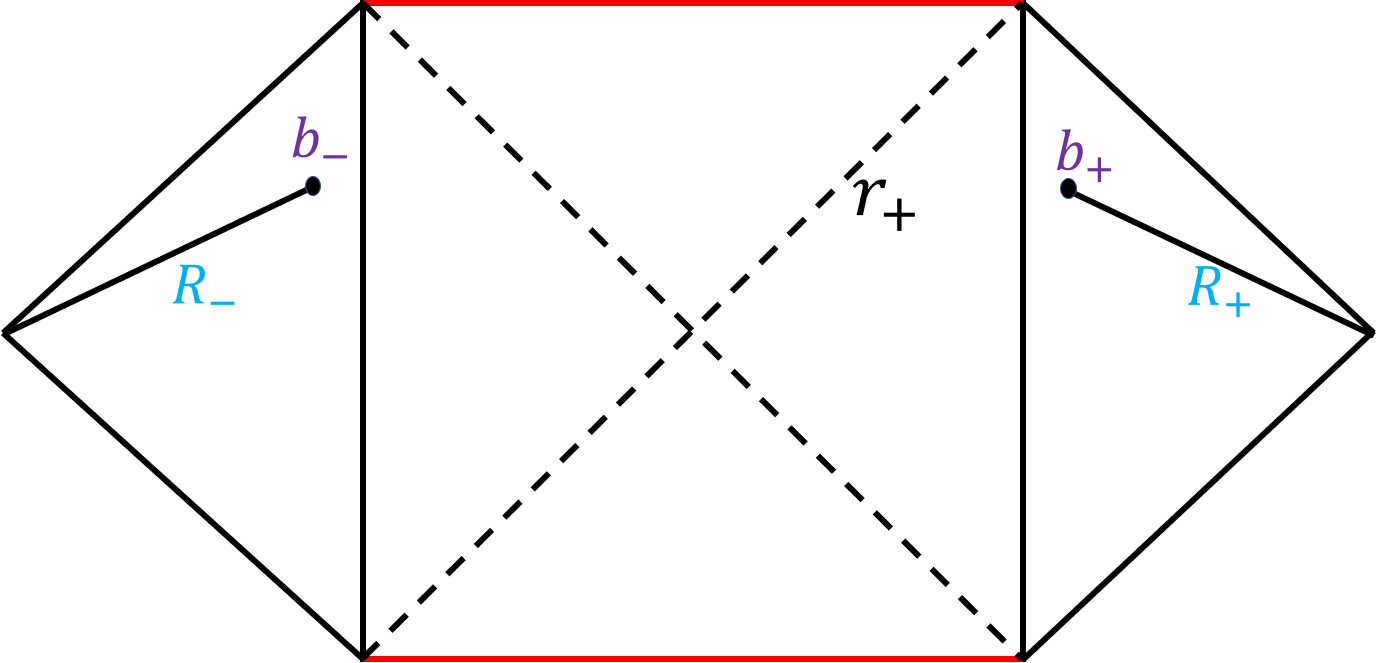}
	\end{center}
	\caption{
		The Penrose diagram of an EGB BH when there is no island.}
	\label{fig: Penrose with island}
\end{figure}

Because there is no island surface initially, Hawking radiation's EE can be estimated using the formula below
\begin{equation} \label{Entropy without island}
    S=\frac{c}{3} \ell(b_+,b_-) \ ,
\end{equation}
where the points $b_\pm = (\pm t_b, b)$  are the radiation region limits in the right and left wedges of the EGB black hole. Defining $\beta$ as inverse Hawking temperature as $\beta=\frac{2 \pi}{k}$. Between the boundary points $b_+$ and $b_-$, the geodesics distance is
\begin{equation} \label{dbb}
    \ell(b_+,b_-) = \sqrt{\frac{4(b^2-r_+^2)[b^2+(b^2-1)\gamma^2]}{(1+\gamma^2)b^2r_+^2} \text{Cosh}^2(r_+t_b))} \ ,
\end{equation} 
putting eq.\eqref{dbb} in eq.\eqref{Entropy without island} we get
\begin{equation}
    S(R)=\left(\frac{c}{6}\right)\text{log} \left[\frac{4(b^2-r_+^2)[b^2+(b^2-1)\gamma^2]}{(1+\gamma^2)b^2r_+^2} \text{Cosh}^2(r_+t))\right] \ ,
\end{equation}  
$c$ is the central charge. At late time i.e., $t \rightarrow \infty$ then $\text{Cosh}(r_+t)\sim e^{r_+t}$, the above equation can be approximated as 
\begin{equation}\label{Sr}
    S(R) \sim \frac{c}{3} r_+ t \ .
\end{equation}
As a result of equation \eqref{S(R)},  we can see that the EE of the Hawking radiation in the absence of an island surface increases linearly with time and becomes infinite at late times, resulting in the EGB black hole's information paradox.
Next, We'll show that at late periods, an island appears, and the Hawking radiation's EE in the presence of an island surface remains constant and dominates after the Page time. The Page curve is obtained by combining the two contributions.
\par 
\subsection{ Entanglement Entropy with Island}
The assumption for the verification of the presence of the island in the EGB BH-Bath System are : 

\begin{figure}[ht]
	\begin{center}
		\includegraphics[scale=0.45]{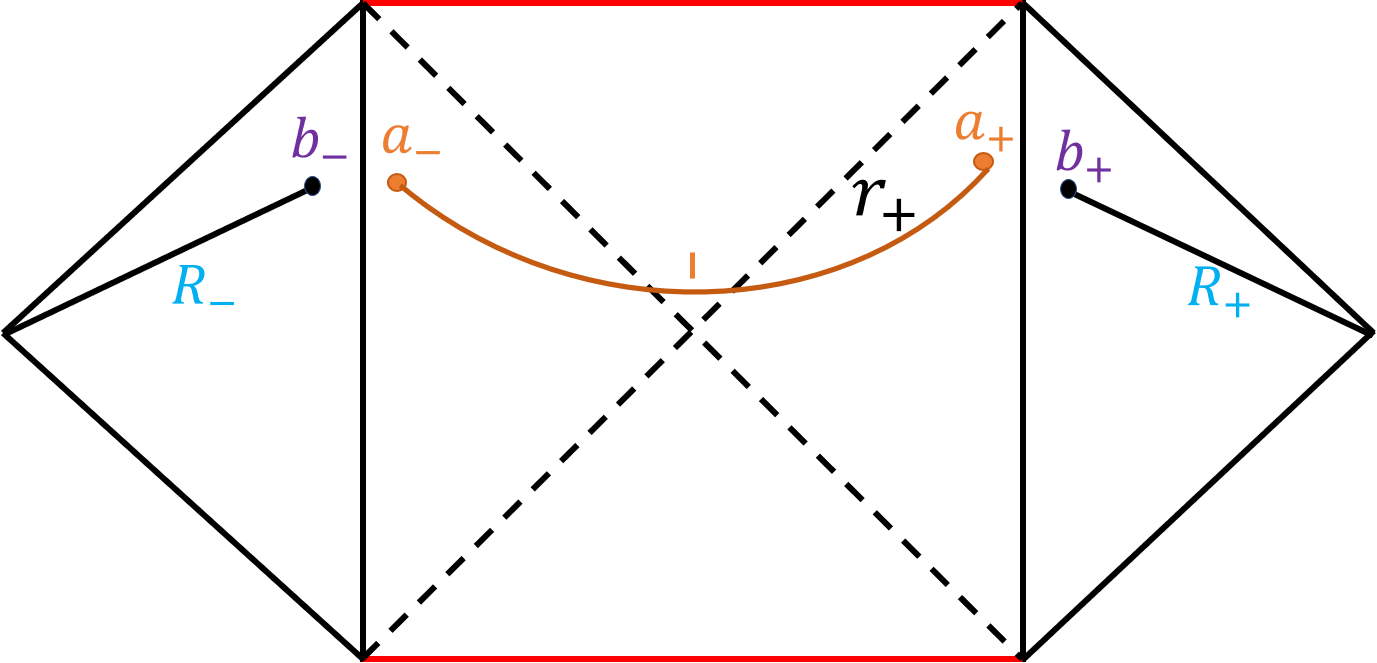}
	\end{center}
	\caption{
		The Penrose diagram of EGB BH when there is an island $I$ indicated in orange with endpoints $a_-$ and $a_+$.}
	\label{fig: Penrose with island}
\end{figure}

 \begin{itemize}
    \item The right wedge is far apart from the left wedge such that the s-wave approximation can be applied.  
    
    \item The quantum matter's von Neumann entropy throughout the entire region for the $R \cup I$ is sufficiently large.
\end{itemize}
Using eq.\eqref{a+b+}, one may determine the von Neumann entropy, one can get the expression for $ S_{\text{generalized}}$. Extremizing the $S_{\text{generalized}}$ with respect to $t_a$ yields i.e., $\frac{\partial S_{\text{generalized}}}{\partial t_a}=0$ results in $t_a=t_b$. By executing this result, one can easily conclude that 
\begin{eqnarray} \label{Sgeneralized}
 S_{\text{generalized}} &=& \frac{\pi a}{G_N} + \frac{c}{6}\text{log} \Bigg[ \frac{1}{r_+^2}\sqrt{\frac{(b^2-r_+^2)[b^2+(b^2-1)\gamma^2]}{b^2(1+\gamma^2)}} \sqrt{\frac{(a^2-r_+^2)[a^2+(a^2-1)\gamma^2]}{a^2(1+\gamma^2)}}  \nonumber \\ 
 && \Bigg(\frac{1}{(b-r_+)(f-r_+)} \Bigg\{ \sqrt{\frac{(b-r_+)}{(b+r_+)}\left[\frac{(b+\gamma r_+)}{(b-\gamma r_+)}\right]^\gamma} \sqrt{\frac{(a-r_+)}{(a+r_+)}\left[\frac{(a+\gamma r_+)}{(a-\gamma r_+)}\right]^\gamma} \nonumber \\
 && \Bigg( (f-r_+)(b+r_+) \left[\frac{b-\gamma r_+}{b+\gamma r_+}\right]^\gamma +  (b-r_+)(f+r_+) \left[\frac{f-\gamma r_+}{f+\gamma r_+}\right]^\gamma \Bigg) - 2  \Bigg \} \Bigg] \nonumber \ .
\end{eqnarray}
The location of QES can be obtained by extremizing the above equation w.r.t $a$

\begin{equation}
    a=r_++ \frac{1}{2r_+}\left(\frac{c G}{6 \pi }\right)^2\left(\frac{b+r_+}{b-r_+}\right) \left(\frac{1+\gamma}{1-\gamma}\right)^\gamma \left(\frac{b-\gamma r_+}{b+\gamma r_+}\right)^\gamma \ .
\end{equation}
Substituting this value in $S_{generalized}$, we have 
\begin{eqnarray} \label{Sgeneralized}
    S_{generalized} &\simeq& \frac{\pi r_+}{ G_N} + \frac{c^2 G_N}{72 \pi^2 }\left[\frac{1+\gamma }{1- \gamma } \right]^{\gamma} + \order{ G_N^2} \nonumber \\
    &\simeq& 2 S_{BH} + \frac{c^2 G_N}{72 \pi^2 }\left[\frac{1+\gamma }{1- \gamma } \right]^{\gamma} + \order{ G_N^2} \ .
\end{eqnarray}
The above expression has some notable characteristics. Aside from the first item $S_{BH}$, the phrase includes universal adjustments to the Hawking entropy of a BH. So, these corrections can be thought of as quantum gravity signals, this is a surprise. The information loss paradox is solved by the above conclusion yielding the proper Page curve and has a lovely mutual information interpretation.
\section{Page Curve} \label{Page time and scrambling time}
 \begin{figure}[ht]
    \centering
    \includegraphics[scale=0.50]{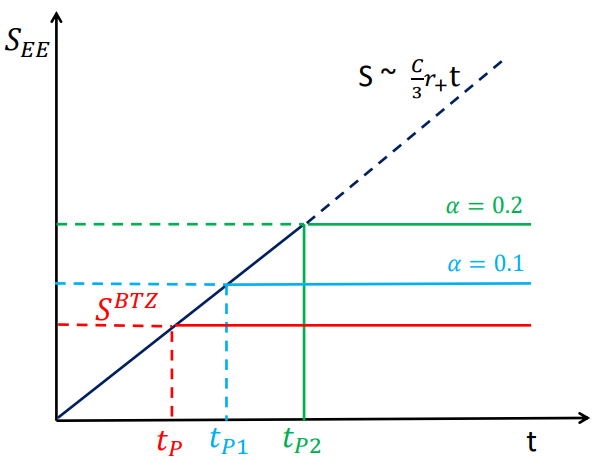}
    \caption{Page curves of an eternal BTZ black hole and for EGB BH for various values of Gauss-Bonnet coupling($\alpha$).}
    \label{Plot}
\end{figure}
\begin{itemize}
    \item \textbf{Page time($t_{\text{page}}$) :} Page time is the point at which Hawking radiation's EE reaches its maximum. The entropy of an eternal black hole does not alter after it.
    By comparing \eqref{Sr} and \eqref{Sgeneralized} 
    \begin{eqnarray}
        t_{\text{page}} &=& \frac{6 }{ c r_+} S^{BH} + \frac{3}{cr_+}\frac{c^2 G_N}{72 \pi^2 }\left[\frac{1+\gamma }{1- \gamma } \right]^{\gamma} \nonumber\\
        &=& \frac{3 \beta}{2 c \pi^2} S_{BH} + \frac{(cG_N)}{96 \pi^4}  \left[\frac{1+\gamma }{1- \gamma } \right]^{\gamma} \ .
    \end{eqnarray}
    In the above expression we have used the expression $\beta=\frac{2 \pi}{T} = \frac{4 \pi^2}{r_+}$.

    \item \textbf{Scrambling time ($t_{\text{scr}}$):} the Hayden-Preskill experiment \cite{Hayden:2007cs} establishes the 
    scrambling time is the smallest time during which information can be retrieved from Hawking radiation. Due to connections between the island's degree of freedom and radiation regions, the scrambling time is also given by the time for information to enter the island in the entanglement wedge reconstruction proposal \cite{Penington:2019npb}. One can compute the Scrambling time by 
     \begin{eqnarray}
     t_{\text{scr}} &\equiv& |r_*(b) -r_*(a)| \nonumber \\
                    &=& \frac{1}{2r_+} \text{log}\left[\left(\frac{b-r_+}{b+r_+}\frac{a+r_+}{a-r_+}\right)\left(\frac{\Gamma b+r_+}{\Gamma b-r_+} \frac{\Gamma a+r_+}{\Gamma a+r_+} \right)^{1/\Gamma}\right] \nonumber \\
                    &\simeq& \frac{\beta}{2 \pi}\text{log}\left(\frac{r_+}{c G_N}\right) + \cdots \nonumber \ .
    \end{eqnarray}
    The higher derivative terms do not affect scrambling time, similar to the BTZ BH's scrambling time.
\end{itemize}
	 

  \section{Results and Discussion} \label{Conclusion}
  In this paper, we have calculated the holographic EE to calculate the Page curves of an evaporating EGB BH in higher derivative terms. In this paper, we have considered $\order{R^2}$ terms in gravitational action and using that, we have plotted the Page curves \ref{Plot}. As a result, when we consider the $\order{R^2}$ terms, initially due to the absence of an island, we obtain that there is linear dependence of time in the EE, resulting in an information paradox for the EGB BH. Late in the game, an island appears, and the Hawking radiation's EE achieves a constant value equal to double the BH's Bekenstein-Hawking entropy, yielding Page curves for fixed values of the Gauss-Bonnet coupling ($\alpha$).  We have also plotted the EE $(S_{\text{EE}})$ with respect to time $(t)$ with fixed parameter $( c=1, G_N=1, r_+=1)$ for BTZ black hole and then $\alpha=0.1$ and $0.2$ for EGB BH. From the figure \ref{Plot}, it is clear that the Page curves change towards later times when Gauss-Bonnet coupling ($\alpha$) rises in this example. We have also studied the page time and the scrambling time in BTZ and the EGB case. For consistency check, for $\alpha \rightarrow 0$, we get a similar expression as in the BTZ case from the EGB case.
  \par

  As we have discussed in the section\label{EGB}, we have considered some specific choices for $\phi(r)$ and $\lambda=0$. It is interesting to investigate what is the expression for the function $f(r)$ if we considered the case with $\lambda\neq 0$. We will come back to these issues later on.


  \section*{Acknwoledgment}
  I am indebted to Prasanta K. Tripathy for a careful manuscript reading.
		\begin{appendices}

\section{Expressions for distances}\label{Distance Expression}
The Expressions for distances are
\begin{eqnarray} \label{a+b+}
 \ell (a_+,b_+) &=& \Bigg[ \frac{1}{r_+^2}\sqrt{\frac{(b^2-r_+^2)[b^2+(b^2-1)\gamma^2]}{b^2(1+\gamma^2)}} \sqrt{\frac{(a^2-r_+^2)[a^2+(a^2-1)\gamma^2]}{a^2(1+\gamma^2)}} \Bigg(\frac{1}{(b-r_+)(f-r_+)} \nonumber \\ 
 && \Bigg\{ \sqrt{\frac{(b-r_+)}{(b+r_+)}\left[\frac{(b+\gamma r_+)}{(b-\gamma r_+)}\right]^\gamma} \sqrt{\frac{(a-r_+)}{(a+r_+)}\left[\frac{(a+\gamma r_+)}{(a-\gamma r_+)}\right]^\gamma} \Bigg( (f-r_+)(b+r_+) \left[\frac{b-\gamma r_+}{b+\gamma r_+}\right]^\gamma \nonumber \\
 && +  (b-r_+)(f+r_+) \left[\frac{f-\gamma r_+}{f+\gamma r_+}\right]^\gamma \Bigg) - 2 \text{Cosh}[r_+(t_a-t_b)] \Bigg \} \Bigg]^{\frac{1}{2}} \ , \\ \hline \hline \nonumber
 \\
 \ell( a_+, b_-) &=& \Bigg[ \frac{1}{r_+^2}\sqrt{\frac{(b^2-r_+^2)[b^2+(b^2-1)\gamma^2]}{b^2(1+\gamma^2)}} \sqrt{\frac{(a^2-r_+^2)[a^2+(a^2-1)\gamma^2]}{a^2(1+\gamma^2)}} \Bigg(\frac{1}{(b-r_+)(f-r_+)} \nonumber \\ 
 && \Bigg\{ \sqrt{\frac{(b-r_+)}{(b+r_+)}\left[\frac{(b+\gamma r_+)}{(b-\gamma r_+)}\right]^\gamma} \sqrt{\frac{(a-r_+)}{(a+r_+)}\left[\frac{(a+\gamma r_+)}{(a-\gamma r_+)}\right]^\gamma} \Bigg( (f-r_+)(b+r_+) \left[\frac{b-\gamma r_+}{b+\gamma r_+}\right]^\gamma \nonumber \\
 && +  (b-r_+)(f+r_+) \left[\frac{f-\gamma r_+}{f+\gamma r_+}\right]^\gamma \Bigg) - 2 \text{Cosh}[r_+(t_a+t_b)] \Bigg \} \Bigg]^{\frac{1}{2}}   = \ell(a_-, b_-) \ , \\ \hline \hline \nonumber
 \\
\ell(b_+,b_-) &=&  \sqrt{\frac{4(b^2-r_+^2)[b^2+(b^2-1)\gamma^2]}{(1+\gamma^2)b^2r_+^2} \text{Cosh}^2(r_+t_b))}  \ . 
\end{eqnarray}
\end{appendices}



\end{document}